\documentclass[sigconf,9pt]{acmart}
\usepackage{enumitem} 
\usepackage{multirow} 

\newtheoremstyle{examplestyle} % name
    {0pt} % Space above
    {0pt} % Space below
    {\itshape} % Body font
    {} % Indent amount
    {\bfseries} % Theorem head font
    {.} % Punctuation after theorem head
    {.1em} % Space after theorem head
    {} % Theorem head spec (can be left empty, meaning 'normal')
\theoremstyle{examplestyle}

\newtheoremstyle{definitionstyle} % name of the style
    {0pt} % Space above
    {0pt} % Space below
    {\itshape} % Body font
    {} % Indent amount
    {\bfseries} % Theorem head font (bold and small caps)
    {.} % Punctuation after theorem head
    {.3em} % Space after theorem head
    {} % Theorem head spec (can be left empty, meaning 'normal')
\theoremstyle{definitionstyle}
\newtheorem{definition}{Definition}
\AtBeginDocument{%
  }

\copyrightyear{2026}
\acmYear{2026}
\setcopyright{cc}
\setcctype{by}
\acmConference[SIGMOD Companion '26]{Companion of the International Conference on Management of Data}{May 31-June 05, 2026}{Bengaluru, India}
\acmBooktitle{Companion of the International Conference on Management of Data (SIGMOD Companion '26), May 31-June 05, 2026, Bengaluru, India}
\acmDOI{10.1145/3788853.3801591}
\acmISBN{979-8-4007-2450-3/2026/05}

\begin{document}

%%
%% The "title" command has an optional parameter,
%% allowing the author to define a "short title" to be used in page headers.
\title{\texttt{CLEAR}: A Knowledge-Centric Vessel Trajectory Analysis Platform}

%%
%% The "author" command and its associated commands are used to define
%% the authors and their affiliations.
%% Of note is the shared affiliation of the first two authors, and the
%% "authornote" and "authornotemark" commands
%% used to denote shared contribution to the research.
\author{Hengyu Liu}
\orcid{0000-0001-6545-7181}
\affiliation{%
  \department{Department of Computer Science}
  \institution{Aalborg University, Denmark}
  \country{}
}
\email{heli@cs.aau.dk}

\author{Tianyi Li}
\authornote{Corresponding author.}
\orcid{0000-0001-5424-6442}
\affiliation{%
  \department{Department of Computer Science}
  \institution{Aalborg University, Denmark}
  \country{}
}
\email{tianyi@cs.aau.dk}

\author{Haoyu Wang}
\orcid{0009-0001-0418-7111}
\affiliation{%
  \department{School of Computer Science and Engineering, Northeastern University}
  \institution{}
  \city{Shenyang, 110819}
  \country{China}
}
\email{haoyu4260@gmail.com}

\author{Kristian Torp}
\orcid{0000-0002-8239-0262}
\author{Yushuai Li}
\orcid{0000-0002-3043-3777}
\affiliation{%
  \department{Department of Computer Science}
  \institution{Aalborg University, Denmark}
  \country{}
}
\email{{torp,yusli}@cs.aau.dk}

\author{Tiancheng Zhang}
\orcid{0000-0001-6902-9299}
\affiliation{%
  \department{School of Computer Science and Engineering, Northeastern University}
  \institution{}
  \city{Shenyang 110819}
  \country{China}
}
\email{tczhang@mail.neu.edu.cn}

\author{Torben Bach Pedersen}
\orcid{0000-0002-1615-777X}
\author{Christian S. Jensen}
\orcid{0000-0002-9697-7670}
\affiliation{%
  \department{Department of Computer Science}
  \institution{Aalborg University, Denmark}
  \country{}
}
\email{{tbp,csj}@cs.aau.dk}
%%
%% By default, the full list of authors will be used in the page
%% headers. Often, this list is too long, and will overlap
%% other information printed in the page headers. This command allows
%% the author to define a more concise list
%% of authors' names for this purpose.
\renewcommand{\shortauthors}{Hengyu et al.}

%%
%% The abstract is a short summary of the work to be presented in the
%% article.
\begin{abstract}
Vessel trajectory data from the Automatic Identification System (AIS) is used widely in maritime analytics. Yet, analysis is difficult for non-expert users due to the incompleteness and complexity of AIS data.
We present \texttt{CLEAR}, a knowledge-\underline{\textbf{c}}entric vesse\underline{\textbf{l}} traj\underline{\textbf{e}}ctory \underline{\textbf{a}}nalysis platfo\underline{\textbf{r}}m that aims to overcome these barriers. 
By leveraging the reasoning and generative capabilities of Large Language Models (LLMs), \texttt{CLEAR} transforms raw AIS data into complete, interpretable, and easily explorable vessel trajectories through a Structured Data-derived Knowledge Graph (SD-KG).
As part of the demo, participants can configure parameters to automatically download and process AIS data, observe how trajectories are completed and annotated, inspect both raw and imputed segments together with their SD-KG evidence, and interactively explore the SD-KG through a dedicated graph viewer, gaining an intuitive and transparent understanding of vessel movements\footnote{See the [\href{https://github.com/hyLiu1994/CLEAR/blob/main/docs/demo.mp4}{demo video}] and our [\href{https://github.com/hyLiu1994/CLEAR}{repository}] containing code and datasets.}.
\end{abstract}

%%
%% The code below is generated by the tool at http://dl.acm.org/ccs.cfm.
%% Please copy and paste the code instead of the example below.
%%


%%
%% Keywords. The author(s) should pick words that accurately describe
%% the work being presented. Separate the keywords with commas.
%% A "teaser" image appears between the author and affiliation
%% information and the body of the document, and typically spans the
%% page.
% \begin{teaserfigure}
%   \includegraphics[width=\textwidth]{sampleteaser}
%   \caption{Seattle Mariners at Spring Training, 2010.}
%   \Description{Enjoying the baseball game from the third-base
%   seats. Ichiro Suzuki preparing to bat.}
%   \label{fig:teaser}
% \end{teaserfigure}

% \received{20 February 2007}
% \received[revised]{12 March 2009}
% \received[accepted]{5 June 2009}

%%
%% This command processes the author and affiliation and title
%% information and builds the first part of the formatted document.
\maketitle

\section{Introduction}
Trajectory data collected from the Automatic Identification System (AIS) captures the movement of vessels~\cite{bank_methodological_2023}. 
Analyzing this data enables analysts to uncover characteristic movement patterns~\cite{li2022evolutionary}, assess operational efficiency~\cite{li2020compression,li2021trace}, and detect potentially unsafe or non-compliant behavior in protected maritime zones~\cite{yang2024harnessing}. 

\begin{figure}[!tbp]
% \vspace{-2mm}
\centering
\includegraphics[width=0.47\textwidth]{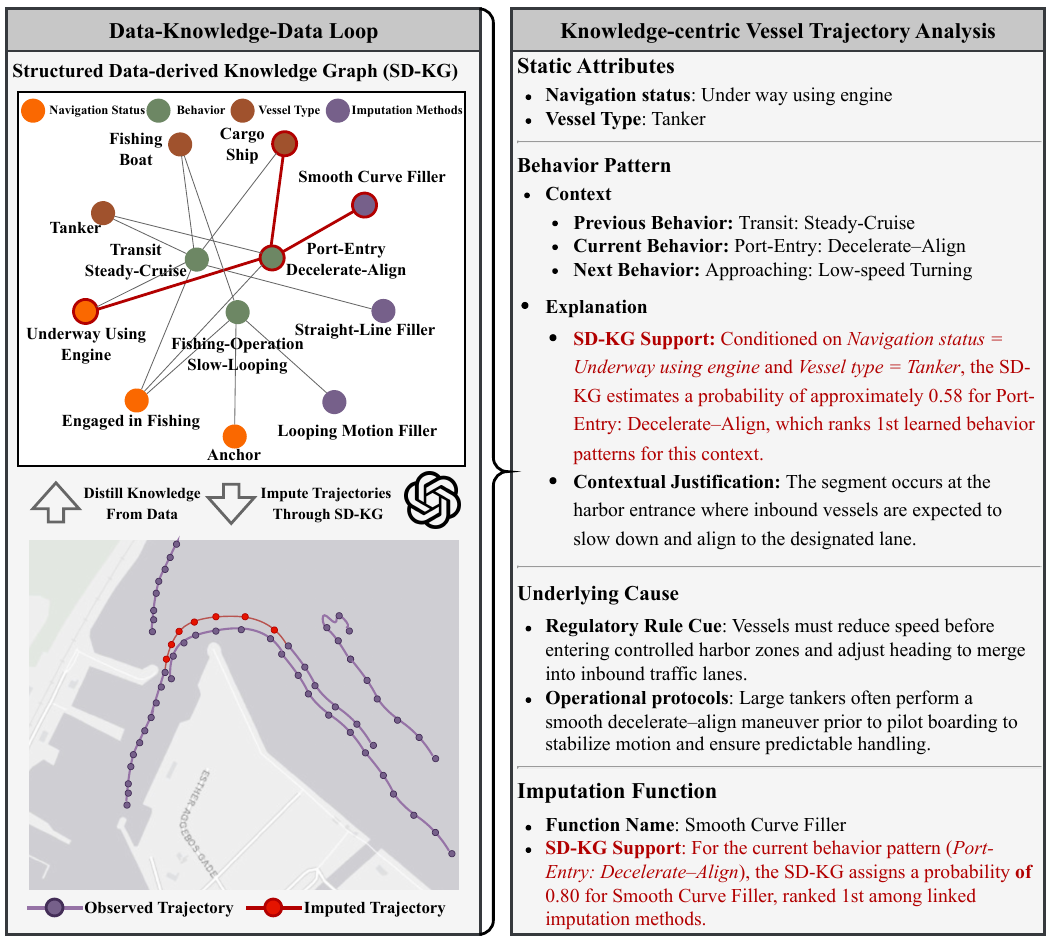}
\vspace{-2mm}
\caption{Illustrative example of \texttt{CLEAR}.}
% : Data-Knowledge-Data loop with SD-KG encoding navigation status, vessel types, behavior patterns and imputation methods (left), and knowledge-centric vessel trajectory analysis with contextual explanation and underlying causes (right)
\vspace{-9mm}
\label{fig:toy_example}
\end{figure}

However, AIS data exhibits two properties that make such analysis challenging in practice. 
First, \textbf{missing data} occurs pervasively: this can be due to communication outages, intentional AIS switch-off, or heterogeneous sampling policies. This in turn can severely distort downstream analysis~\cite{mh-gin-pvldb-2025,galdelli2025data,jeung2008discovery}. 
Second, AIS data is \textbf{heterogeneous, and semantically complex}: meaningful interpretation often requires combining kinematic signals with static vessel attributes, regulations, and contextual information such as traffic separation schemes or port layouts~\cite{bank_methodological_2023,yang2024harnessing}. As a result, to be able to perform effective trajectory analyses typically requires analysts to possess both expertise in spatio-temporal databases (such as PostGIS~\cite{PostGIS_2026} and MobilityDB~\cite{zimanyi2020mobilitydb}) and maritime domain knowledge.
Together, these characteristics make \textbf{AIS trajectory analysis particularly challenging for non-expert users}.

While several trajectory analysis tools exist, they provide limited support for non-expert users: maritime analytics platforms (e.g., MarineTraffic~\cite{MarineTraffic_2026}, and VesselFinder~\cite{VesselFinder_2026}) offer intuitive visualization but lack deeper behavioral interpretation and operate only on raw incomplete data; trajectory analytics tools (e.g., MovingPandas~\cite{graser2019movingpandas}) require programming skills and provide no domain-specific knowledge or explanations for observed movement patterns.

We present \texttt{CLEAR}, a knowledge-\underline{\textbf{c}}entric vesse\underline{\textbf{l}} traj\underline{\textbf{e}}ctory \underline{\textbf{a}}nalysis platfo\underline{\textbf{r}}m that makes AIS trajectory analysis accessible to non-expert users by transforming raw AIS data into complete, interpretable, and easily explorable vessel trajectories through a Structured Data-derived Knowledge Graph (SD-KG). 
As shown in Figure~\ref{fig:toy_example}, \texttt{CLEAR} operates through two tightly coupled stages. 
At its core, \texttt{CLEAR} implements the \emph{data--knowledge--data loop} introduced in the VISTA framework~\cite{liu2026vista}, which uses a Structured Data-derived Knowledge Graph (SD-KG) to distill structured maritime knowledge from AIS data and then reuse it to guide trajectory imputation with evidence-backed explanations.
On top of this loop, \texttt{CLEAR} enables \emph{knowledge-centric analyses}: it exposes segment-level evidence for each (raw or imputed) trajectory segment and enables exploration of the SD-KG. 

% In the data–knowledge–data loop, \texttt{CLEAR} leverages the semantic reasoning and generative capabilities of Large Language Models (LLMs) to first distill structured knowledge from AIS data and then apply this knowledge back to guide trajectory imputation. 
% In the knowledge-centric analysis stage, \texttt{CLEAR} exposes the relevant SD-KG evidence at the segment~level and support exploration in SD-KG. 

\noindent\textbf{Data–Knowledge–Data Loop.}
As illustrated to the left in Figure~\ref{fig:toy_example}, \texttt{CLEAR} first distills knowledge from the observed trajectories (purple) to construct the SD-KG. 
In this example, the SD-KG integrates four types of nodes, \emph{vessel types} (brown), \emph{navigation statuses} (orange), \emph{behavior patterns} (green), and \emph{executable imputation methods} (purple), with edge weights encoding how frequently these elements co-occur in segments. 
Using this graph, \texttt{CLEAR} then imputes missing trajectory parts: for the gap highlighted in red, it queries the SD-KG with the local context and infers that the segment most likely corresponds to a \emph{Port-Entry: Decelerate–Align} maneuver performed by a \emph{Cargo} vessel in the status \emph{Underway using engine}. 
Based on these inferred attributes and behavior pattern, the SD-KG ranks \emph{Smooth Curve Filler} as the most suitable method, yielding the red imputed segment shown on the map. 
This process is \emph{automatic}, \emph{transparent}, and \emph{interpretable}---key properties for trustworthy AI systems in safety-critical domains such as 6G-enabled maritime networks~\cite{xylouris20246g}, where decisions must be accountable and auditable.

\noindent\textbf{Knowledge-centric Vessel Trajectory Analysis.}
\texttt{CLEAR} provides a \emph{knowledge-centric analysis} for every segment. As illustrated to the right in Figure~\ref{fig:toy_example}, for the imputed segment highlighted in red, the system presents its associated \emph{static attributes}, the inferred \emph{behavior context} (previous, current, and next behaviors), and the \emph{explanation} supporting these estimates.  
The information highlighted in red is shown only for imputed segments, providing insight into why the system inferred a particular \emph{behavior pattern} and selected a specific \emph{imputation method}.  
This analysis is designed to be accessible to \emph{non-experts}: users without maritime expertise can interpret segment behavior, and deepen their understanding by exploring the SD-KG.

The demo allows conference attendees to experience \texttt{CLEAR} in action. Participants can specify parameters to automatically download AIS data from the data source for automated processing, view completed trajectories produced by \texttt{CLEAR}, and explore how the system augments raw trajectories with imputed segments and segment-level knowledge. Participants may inspect both raw and imputed segments, examine their associated patterns and static attributes, and explore the SD-KG through a dedicated graph viewer that exposes the semantics and relationships underlying \texttt{CLEAR}'s outputs. Together, these components allow participants to follow \texttt{CLEAR}'s workflow---from the knowledge used to enrich trajectories, to the resulting imputed tracks, and ultimately to downstream trajectory analysis---in an intuitive and transparent manner.

\section{Background}
Before detailing \texttt{CLEAR}, we present the AIS data and knowledge structures it builds on. 
Definition~\ref{def-1} specifies the attributes of an AIS record, and Definition~\ref{def-2} introduces the Structured Data-derived Knowledge Graph (SD-KG), which provides a structured representation of the knowledge abstracted from AIS data in the form of static attributes, behavior patterns, and imputation methods.

\begin{definition}[\textbf{AIS Record}] \label{def-1}
An AIS record captures a vessel's state at a specific time and location, comprising: (i)~\textbf{spatio-temporal attributes} (vessel ID, longitude, latitude, timestamp); (ii)~\textbf{kinematic attributes} (heading, course, speed); (iii)~\textbf{status-related attributes} (navigation status, cargo type, draught); and (iv)~\textbf{static attributes} (vessel dimensions and type).
\end{definition}

\begin{definition}[\textbf{Structured Data-derived Knowledge Graph}] \label{def-2}
An SD-KG is a weighted tripartite graph $\mathcal{G}_d = (\mathcal{V}_s \cup \mathcal{V}_b \cup \mathcal{V}_f,\; \mathcal{E}_{\mathit{sb}} \cup \mathcal{E}_{\mathit{bf}})$ distilled from AIS data, whose three node types directly support \texttt{CLEAR}'s demonstration scenarios:
\begin{itemize}[itemsep=1pt, leftmargin=12pt]
    \item \textbf{Static Attribute Nodes} $\mathcal{V}_s$: vessel types, navigation statuses, and spatial contexts (e.g., ports, shipping lanes), which serve as entry points for filtering and querying in Scenarios~I and~II.
    \item \textbf{Behavior Pattern Nodes} $\mathcal{V}_b$: characteristic movement patterns (e.g., \emph{Port-Entry: Decelerate--Align}) that describe speed, course, heading, intent, and duration, enabling the segment-level explanations shown in Scenario~I.
    \item \textbf{Imputation Method Nodes} $\mathcal{V}_f$: executable imputation functions with descriptions, whose selection rationale is exposed in the knowledge-centric analysis of Scenario~I.
\end{itemize}
Edge weights in $\mathcal{E}_{\mathit{sb}}$ and $\mathcal{E}_{\mathit{bf}}$ encode co-occurrence frequencies, allowing \texttt{CLEAR} to rank candidate behaviors and methods; users can inspect these weights through the graph explorer in Scenario~II.
\end{definition}

\section{\texttt{CLEAR} Overview}
Unlike existing maritime trajectory platforms such as PortVIS~\cite{zhang2025portvis}, which focus on visualization and basic imputation of port-level trajectories, \texttt{CLEAR} provides \emph{knowledge-grounded} analysis where every imputation decision is traceable to structured evidence in the SD-KG and accompanied by human-readable explanations.

\subsection{Data–Knowledge–Data Loop}
\begin{figure}[!tbp]
\centering
\includegraphics[width=1\linewidth]{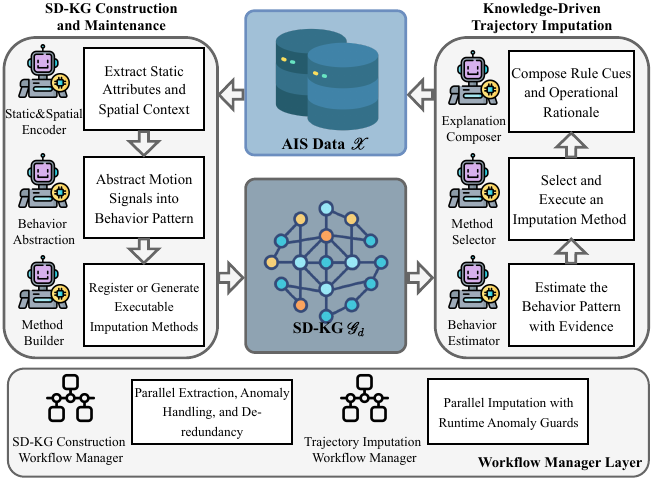}
\vspace{-4mm}
\caption{Overview of the data-knowledge-data loop.}
\label{fig:overview of loop}
\vspace{-4mm}
\end{figure}

\texttt{CLEAR} builds on \texttt{VISTA}~\cite{liu2026vista}, which follows a \emph{data--knowledge--data} loop shown in Figure~\ref{fig:overview of loop} to continuously distill knowledge from AIS data and feed it back to trajectory imputation via the SD-KG. The loop consists of two coupled pipelines, \emph{SD-KG Construction and Maintenance} to the left and \emph{Knowledge-Driven Trajectory Imputation} to the right, orchestrated by a workflow manager layer.

In the SD-KG construction pipeline, \texttt{CLEAR} takes AIS data $\mathcal{X}$ as input and, for each complete trajectory segment, (i) uses a \emph{Static\&Spatial Encoder} to extract static attributes and spatial context, 
(ii) performs \emph{Behavior Abstraction} to compress motion signals into behavior patterns, and 
(iii) invokes a \emph{Method Builder} to register or generate executable imputation methods. 
These LLM-based components populate and update the SD-KG $\mathcal{G}_d$ with static-attribute, behavior-pattern, and imputation-method nodes, while edge weights record how frequently they co-occur or succeed on similar segments. 
In the trajectory imputation pipeline, incomplete trajectories query the SD-KG for knowledge that can be pushed back to the data. 
A \emph{Behavior Estimator} first uses SD-KG statistics and context to estimate the most plausible behavior pattern for a gap, 
after which a \emph{Method Selector} chooses an appropriate imputation method and executes it to reconstruct the missing segment. 
Finally, an \emph{Explanation Composer} summarizes rule cues, operational rationale, and SD-KG evidence for the chosen pattern and method. 

To scale to large AIS collections, both pipelines are orchestrated by the workflow manager layer that parallelizes execution across trajectories and provides de-redundancy, anomaly detection, and anomaly guards during large-scale extraction and imputation.

% Together, these two pipelines close the data–knowledge–data loop. AIS data are converted into structured knowledge in the SD-KG, which is in turn applied to clean and complete AIS trajectories.

\subsection{Knowledge-centric Trajectory Analysis}
\begin{figure}[!tbp]
\centering
\includegraphics[width=1\linewidth]{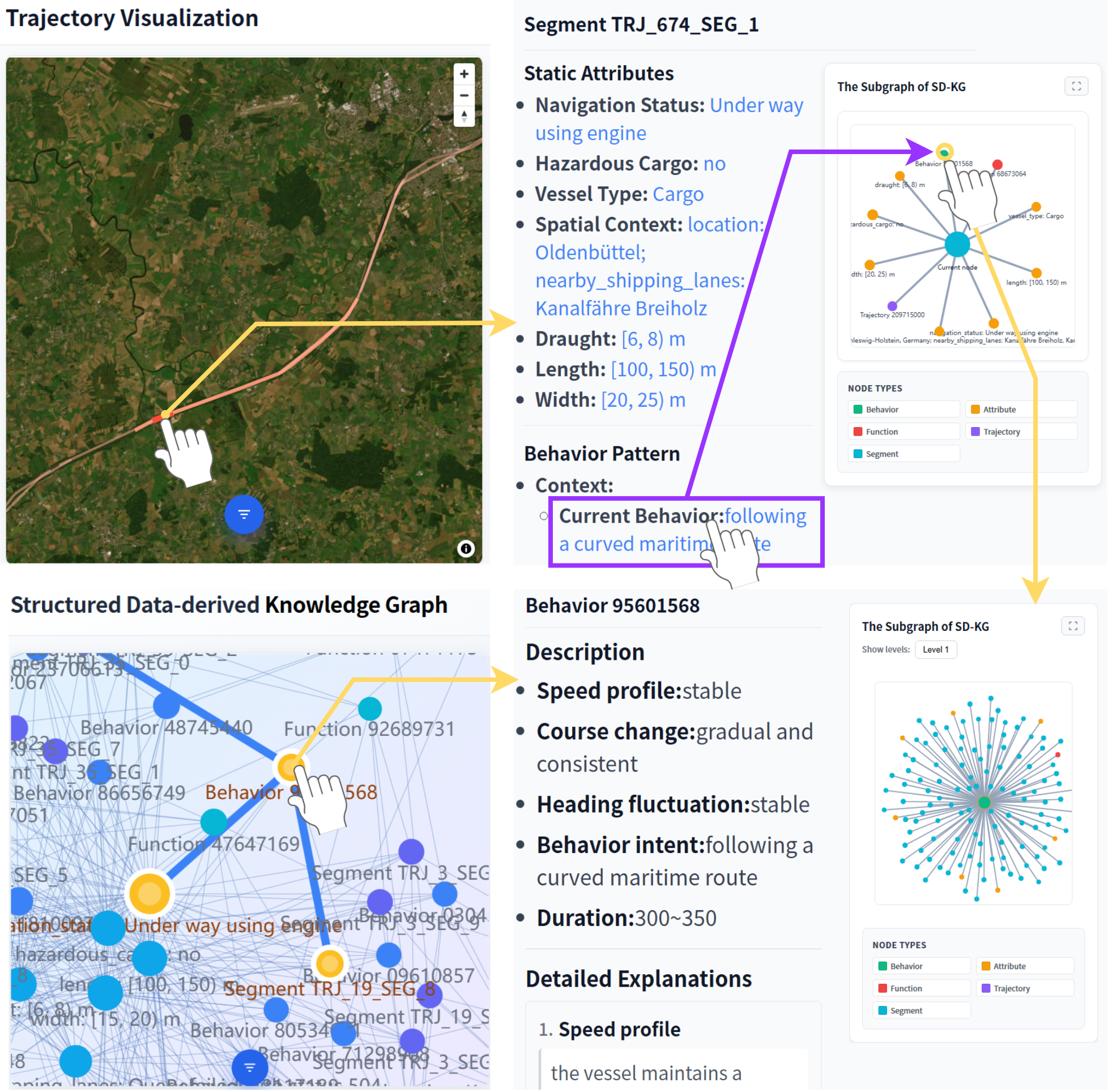}
\vspace{-5mm}
\caption{Overview of knowledge-centric analysis.}
\label{fig:overview of analysis}
\vspace{-5mm}
\end{figure}

As illustrated in Figure~\ref{fig:overview of analysis}, \texttt{CLEAR} offers two core functions, \textbf{Trajectory Analysis} and \textbf{SD-KG Exploration}, that enable non-expert users to
(i) interpret individual trajectories through static attributes, inferred behavioral patterns, and rich contextual information; and
(ii) explore the underlying SD-KG to examine the relationships among behavioral patterns and static attributes, thereby gaining a deeper understanding of the maritime domain.

%Segment-level analysis on the map.
\noindent\textbf{Trajectory Analysis via Map Interaction.}
As shown at the top of Figure~\ref{fig:overview of analysis}, 
\texttt{CLEAR} allows users to click any segment on the map to see a segment-level analysis report. 
The report presents an interactive knowledge-centric analysis, which details the segment's static attributes, inferred behavior context, and explanatory reasoning for behavioral and imputation estimates, alongside a SD-KG subgraph for knowledge exploration. 
These enable non-expert users to intuitively grasp the behavioral semantics of individual trajectory segments and perform rapid pattern comparison.

%SD-KG guided knowledge exploration.
\noindent\textbf{SD-KG Exploration.}
As shown at the upper-right and lower-left of Figure~\ref{fig:overview of analysis}, \texttt{CLEAR} supports two exploration modes, allowing users to explore the SD-KG deeply.
%SD-KG Exploration via reciprocal links between reports. 
First, the SD-KG is explored through links between analysis reports. Within a report, users can view specific information of the current node and explore related nodes, accessing the analysis reports for those nodes. This arrangement connects knowledge across reports and integrates conclusions with evidence, enabling seamless exploration of evidence chains.
%SD-KG Exploration via Graph Viewer.
Second, the SD-KG is explored through a graph viewer. The viewer presents nodes  visually with their direct connections, allowing users to clearly observe relationships between nodes and progressively explore related nodes, thereby gaining an intuitive understanding of each node's context and connections within the~SD-KG.

\begin{figure*}[!tbp]
\centering
\includegraphics[width=1\linewidth]{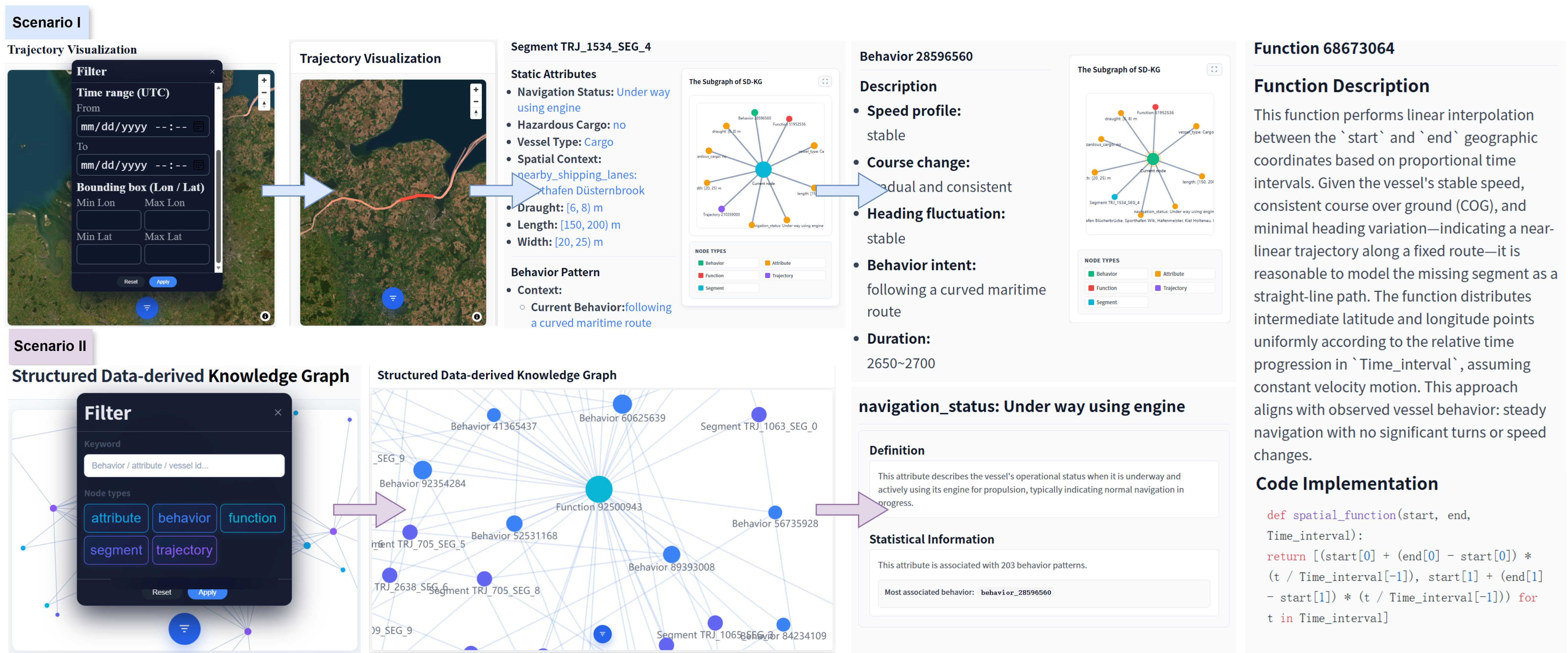}
\vspace{-6mm}
\caption{Scenarios I and II of \texttt{CLEAR}.}
\label{fig:Scenarios I and II}
\vspace{-2mm}
\end{figure*}

\begin{figure}[!tbp]
\centering
\includegraphics[width=1\linewidth]{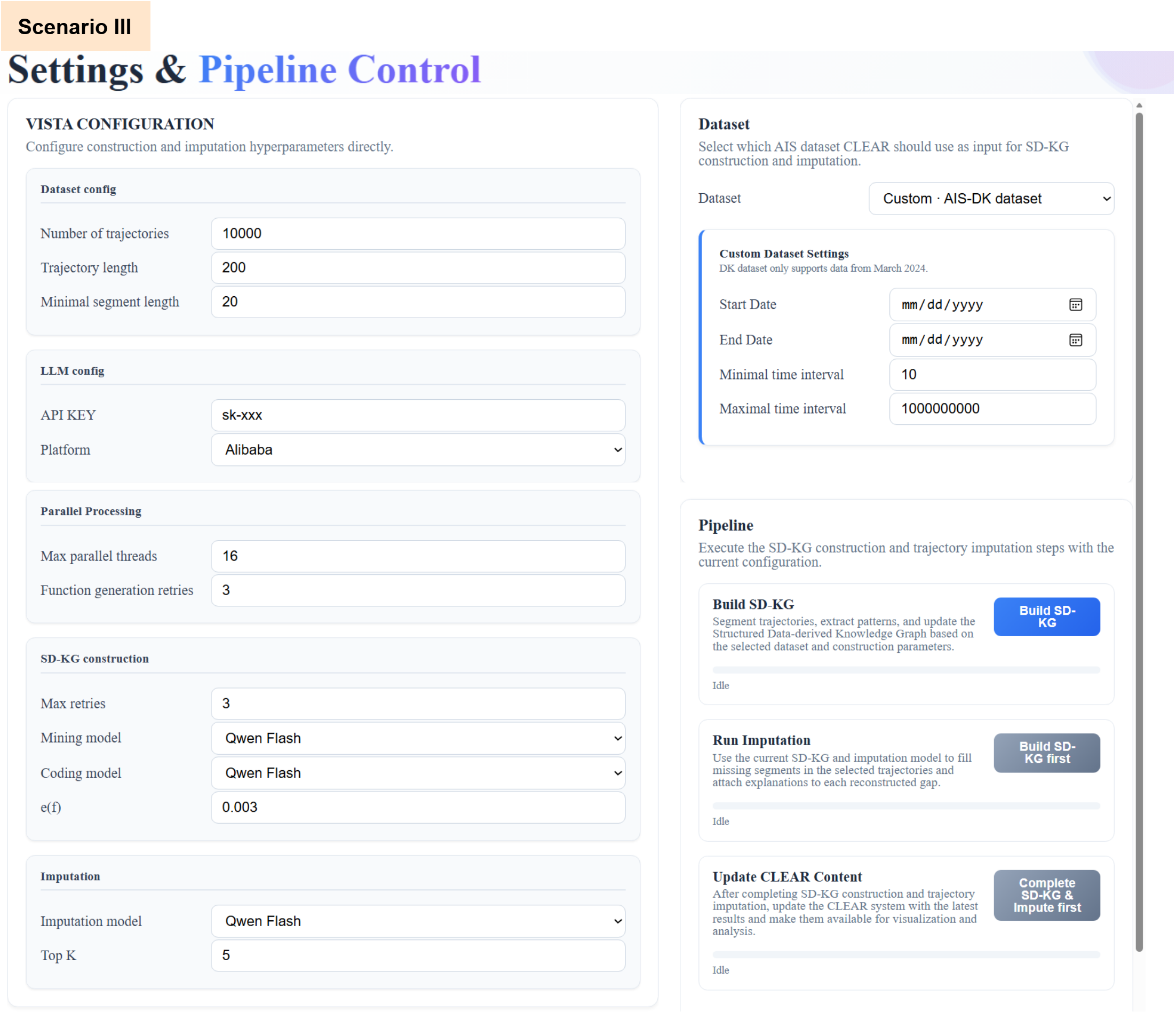}
\vspace{-5mm}
\caption{Scenario III of \texttt{CLEAR}.}
\label{fig:Scenarios III}
\vspace{-5mm}
\end{figure}

\section{Demo Scenarios}
\noindent \textbf{Scenario I: Knowledge-centric Trajectory Analysis.} As shown at the top of Figure~\ref{fig:Scenarios I and II}, participants can filter trajectories on the map based on attributes such as MMSI, time range, or geographic region, and they can click on a trajectory segment to open its knowledge-centric analysis report. The report includes behavior patterns, static vessel attributes, imputation methods applied to the segment, and an SD-KG subgraph showing nodes related to that segment. Users can click on any field in the report to highlight the corresponding nodes in the SD-KG, and clicking on these nodes opens their associated analysis reports, providing an intuitive view of the relationship between analysis results and the underlying knowledge structure.

\noindent \textbf{Scenario II: Exploring the SD-KG.} As shown at the bottom of Figure~\ref{fig:Scenarios I and II}, users can explore the SD-KG via a dedicated graph viewer to examine its overall structure. They can filter nodes by type or keyword, they can can drag any node to highlight all connected edges within the SD-KG, and they can zoom in to view node labels or click on nodes to view their knowledge-centric analysis reports along with SD-KG subgraphs of related nodes. Additionally, nodes in the SD-KG subgraph support click-to-navigate functionality, allowing navigation from a graph node to its corresponding analysis report. This interaction allows users to quickly map abstract knowledge to concrete trajectory instances.

\noindent \textbf{Scenario III: Data Source Selection and Processing Configuration.} As shown in Figure~\ref{fig:Scenarios III}, participants can select an AIS data source and can configure dataset parameters including date range and time intervals. The system automatically downloads the corresponding AIS data based on these selections. Participants can then configure \texttt{VISTA} processing options such as the maximum trajectory segment duration. These parameters jointly control the trade-off between SD-KG richness and processing cost. Based on the specified parameters, the system performs SD-KG construction and trajectory imputation, enabling participants to efficiently progress from data source selection to trajectory analysis.

In future, we plan to extend \texttt{CLEAR} from offline to online, enabling real-time, privacy-preserving, and explainable trajectory analysis.

\begin{acks}
This research was supported in part by the European Union's Horizon Europe projects MobiSpaces (grant agreement no. 101070279) and 6G-XCEL (grant agreement no. 101139194), and the National Natural Science Foundation of China (No. 62272093).
\end{acks}

%%
%% The next two lines define the bibliography style to be used, and
%% the bibliography file.
\bibliographystyle{ACM-Reference-Format}
\bibliography{sample-base}
%%
%% If your work has an appendix, this is the place to put it.

\end{document}